\documentclass[graybox]{svmult}

\usepackage{mathptmx}       % selects Times Roman as basic font
\usepackage{helvet}         % selects Helvetica as sans-serif font
\usepackage{courier}        % selects Courier as typewriter font
\usepackage{type1cm}        % activate if the above 3 fonts are
\usepackage{makeidx}         % allows index generation
\usepackage{graphicx}        % standard LaTeX graphics tool
\usepackage{multicol}        % used for the two-column index
\usepackage[bottom]{footmisc}% places footnotes at page bottom

\usepackage{amssymb}
\usepackage{amsmath}
\usepackage{epsfig}
\usepackage{subfigure}
\usepackage{mathrsfs}

\newcommand{\be}{\begin{equation}}
\newcommand{\ee}{\end{equation}}
\newcommand{\bea}{\begin{eqnarray}}
\newcommand{\eea}{\end{eqnarray}}

\makeindex            

\begin{document}

\title*{Decay of the Cosmic Vacuum Energy}

\author{Timothy Clifton and John D. Barrow}
\institute{Timothy Clifton \at School of Physics \& Astronomy, Queen Mary University of London, Mile End Road, London E1 4NS, U.K., \email{t.clifton@qmul.ac.uk}
\and John D. Barrow \at DAMTP, Centre for Mathematical Sciences, Wilberforce Road, Cambridge CB3 0WA, University of Cambridge, U.K., \email{j.d.barrow@damtp.cam.ac.uk}}

\maketitle

\abstract{In his 2005 review, {\it Gravity and the Thermodynamics of Horizons}, Paddy suggested that a vacuum in thermal equilibrium with a bath of radiation should have a gradually diminishing energy. We work through the consequences of this scenario, and find that a coupling between the vacuum and a bath of black-body radiation at the temperature of the horizon requires the Hubble rate, $H$, to approach the same type of evolution as in the \textquotedblleft intermediate inflation\textquotedblright\ scenario,
with $H\propto t^{-1/3}$, rather than as a constant. We show that such behaviour does not conflict with observations when the vacuum energy is described by a slowly-rolling scalar field, and when the fluctuations in the scalar field are treated as in the ``warm inflation'' scenario. It does, however, change the asymptotic states of the universe. We find that the existence of the radiation introduces a curvature singularity at early times, where the energy densities in both the radiation and the vacuum diverge. Furthermore, we show that the introduction of an additional non-interacting perfect fluid into the space-time reveals that radiation can dominate over dust at late times, in contrast to what occurs in the standard cosmological model. Such a coupling can also lead to a negative vacuum energy becoming positive.}

\newpage

\section{Introduction}

\label{sec1}

Inflationary cosmology is based on the hypothesis of a period of accelerated
expansion in the very early history of the universe. This surge in the
expansion solves the horizon problem, and is generically expected to drive
the observable curvature of space, and any expansion or curvature
anisotropies, to unobservably small values today. In addition, inflationary
cosmology provides a natural mechanism for creating the seeds of
structure formation, from tiny quantum mechanical fluctuations. Such
fluctuations are a manifestation of the thermal nature of quantum fields in
curved spaces. But the existence of a thermal space also implies the
existence of a bath of radiation \cite{therm1,therm2,therm3}. In this paper we will consider the
gravitational consequences of this radiation on the large-scale expansion of the
universe, as well as on the observables that emerge from a period of inflation, as suggested by Paddy in his review \cite{pad}. We will work in Planck units throughout, such that $G=c=\hbar=k_B=1$.

The energy density of radiation with a black-body spectrum, at temperature $T$, is given in Planck
units as 
\begin{equation}
\rho _{\mathrm{r}}=4\sigma T^{4} \, ,
%=\frac{g_{\ast }H^{4}}{480\pi ^{2}}\,,
\label{bb}
\end{equation}%
where $\sigma =g_{\ast }\pi ^{2}/120$ is the Stefan-Boltzmann constant, 
%$H$ is the Hubble expansion rate, 
and $g_{\ast }$ is the effective number of relativistic degrees of freedom. In a Friedmann-Lema\^{\i}tre-Robertson-Walker (FLRW) geometry, the
energy-conservation equation for this radiation fluid is given by 
\begin{equation}
\dot{\rho}_{\mathrm{r}}+4 H \rho _{\mathrm{r}}=Q\,,  \label{dr}
\end{equation}%
%where $a=a(t)$ is the scale factor of the universe ($H=\dot{a}/a$), and
where $Q=Q(t)$ is an energy exchange term that is required in order for Eq. (%
\ref{bb}) to be satisfied at all times, and overdots denote derivatives with
respect to the comoving proper time, $t$. The $Q$ term parameterizes the
energy flow into the radiation field, and thermalization is assumed to be
instantaneous.

Energy-momentum conservation now requires that $T_{\phantom{\mu
\nu};\nu }^{\mu \nu }=0$, where $T_{\mu \nu }$ is the total energy-momentum
tensor of all matter fields in the space-time. If we are considering a
space-time that contains effectively just radiation ($r$) and vacuum ($v$)
energy, with $T_{\mu \nu }=T_{\mu \nu }^{\mathrm{r}}+T_{\mu \nu }^{\mathrm{v}%
}$, then we must therefore also have 
\begin{equation}
\dot{\rho}_{\mathrm{v}}=-Q\,,  \label{dv}
\end{equation}%
where $\rho _{\mathrm{v}}=-p_{\mathrm{v}}$ is the energy density of the
vacuum, and $p_{v}$ is its pressure. This equation shows that the vacuum energy
density must be decaying, and is the cosmological counterpart of the
requirement that radiating black holes must reduce in mass, in order for the
total energy-momentum in the space-time to be conserved \cite{BL}.

The equations above, together with the Friedmann equation, 
$H^{2}=\frac{8\pi }{3}(\rho _{\mathrm{r}}+\rho _{\mathrm{v}}) \, ,$
can be used to write
%combined to show that the energy-exchange term, $Q$, is given by 
%\[Q=\frac{g_{\ast }}{120\pi ^{2}}H^{3}(\dot{H}+H^{2}).\]
%This in turn implies 
\begin{equation}
\dot{H}+\frac{g_{\ast }}{90\pi } (2\pi T)^4=0\, .  \label{dH1}
\end{equation}%
If we can find an expression for $T$ as a function of $H$, then we have a differential equation that can be solved to find the rate of expansion.

In this study we will assume that the radiation is in thermal equilibrium with the vacuum, so that $T$ in Eq. (\ref{dH1}) is given by the usual semi-classical expression for the temperature of space:
\begin{equation} \label{T}
T = \frac{\vert \kappa \vert}{2 \pi} = \left\vert \frac{H}{2 \pi} \left(1 + \frac{\dot{H}}{2 H^2} \right)  \right\vert \, ,
\end{equation}
where $\kappa$ is the surface gravity of the apparent horizon. The expression after the second equality is found by evaluating the surface gravity of the horizon in a spatially flat FLRW geometry \cite{cai}. 

\section{Background Evolution}

\begin{figure}[t!]
\begin{center}
\includegraphics[scale=1.5]{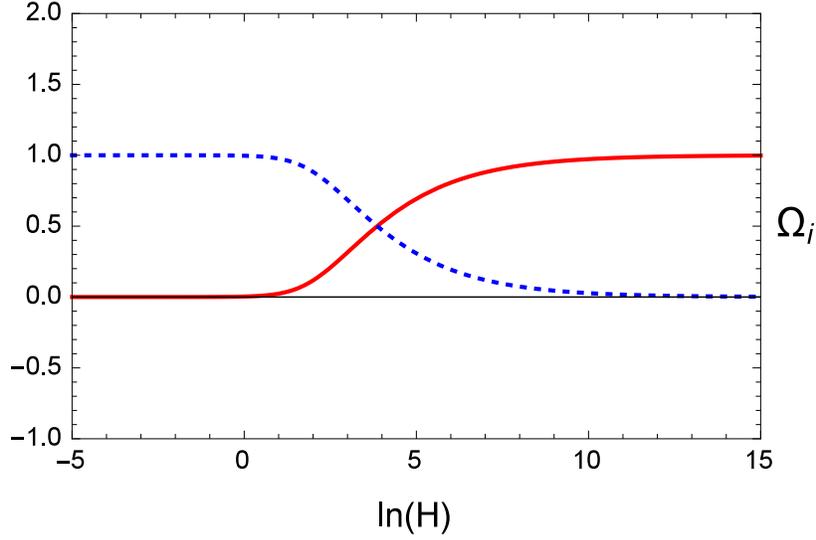}
\end{center}
\caption{The evolution of the density parameters, $\Omega_i=8 \pi \rho_i/3 H^2$, for the radiation field (red solid line) and the vacuum field (blue dotted line). We have chosen $g_{\ast}=2$ in order to produce this plot.}
\label{fig1}
\end{figure}

It can immediately be seen from Eqs. (\ref{dH1}) and (\ref{T}) that $\dot{H} \leq 0$, and that $\dot{H} = 0$ if and only if $H=0$. This shows that $H$ is always decreasing, and that (in an initially expanding universe) it is bounded from below by zero. We can therefore use $H$ as a proxy for time. Figure \ref{fig1} shows the evolution of the energy density in both the radiation and vacuum fields as a function of $H$, as the Universe expands. We have chosen to display this information in terms of the density parameters, defined as $\Omega_i \equiv 8\pi  \rho_i/ 3 H^2$.

At early times, when $H\rightarrow \infty$, it can be seen that $\dot{H} \rightarrow -2 H^2$. Using $H=\dot{a}/a$ this can be shown to correspond to a scale factor of the form
\begin{equation} \label{rad}
a(t) \propto t^{\frac{1}{2}} \, .
\end{equation}
This is the same as that which occurs in a standard radiation-dominated FLRW cosmology. It can be seen from Eq. (\ref{dH1}) that in this limit $T \propto H^{\frac{1}{2}} \propto 1/a$, and from Eq. (\ref{bb}) that $\rho_r \propto H^2 \propto 1/a^4$. These are again exactly the forms of these expressions that one would expect from a radiation filled universe without energy exchange. However, while $\Omega_v \rightarrow 0$ at early times, it can be seen that $\rho_v \rightarrow \infty$. That is, we find that a coupling to radiation can reduce the magnitude of the vacuum energy, even if it is initially very large.

At late-times, on the other hand, we have $H\rightarrow 0$, and hence $\dot{H} /H^2 \rightarrow 0$. In this limit the leading-order contribution to the temperature takes the form of the Gibbons-Hawking value, $T=H/2 \pi$ \cite{GH}, and Eq. (\ref{dH1}) becomes $\dot{H}+{g_{\ast } H^4}/{90\pi }=0$. This leads to a scale factor that evolves as
\begin{equation}
a(t)\propto \exp \left\{ \frac{3}{2}\left( \frac{30\pi }{g_{\ast }}\right) ^{%
\frac{1}{3}}(t-t_{0})^{\frac{2}{3}}\right\} \,,  \label{ii}
\end{equation}%
where $t_{0}=$ constant. This type of expansion is of a type known as
\textquotedblleft intermediate inflation\textquotedblright \cite{inter1a},
which generally has $a(t)\propto \exp \{At^{n}\},$ where $A>0$ and $0<n<1$ are  constants. Intermediate inflation has
been studied by several authors \cite{inter2b}, \cite{mus}, \cite{star}, and is known to
arise in rainbow gravity theories \cite{BM}. 

The particular form of intermediate inflation with $a(t)\propto \exp
\{\lambda t^{2/3}\}$ is special. When generated from a minimally
coupled scalar field in a suitably chosen potential, it is the only form of
intermediate inflation (other than perfect de Sitter) that gives an exact
Harrison-Zeldovich spectrum of first-order density perturbations. Unlike
standard slow-roll scenarios, however, it is also known that this type of
intermediate inflation can produce large amounts of gravitational radiation 
\cite{inter1a, inter2b, inter1b, inter2a}.

Eq. (\ref{ii}) is a
significant departure from the usual exponential expansion, and occurs even
though the energy density of radiation may be small. This can be attributed
to the dual requirements of an almost constant density of radiation, as well
as the exponential dilution of that radiation with inflationary expansion.
Therefore, the vacuum energy must constantly replace the quickly dissipating
radiation, and even though the amount of radiation required at any given
time may be small, it must effectively be replenished at every moment of
time.

\section{Energy Exchange in the Presence of a Non-Interacting Fluid}

If we also include in the universe a non-interacting perfect fluid, with equation of state $$p=(\gamma -1)\rho, $$ then the Friedmann equation becomes 
\begin{equation} \label{fried}
H^{2}=\frac{8\pi }{3}\left( \rho _{r}+\rho _{v}+\rho \right) \, ,
\end{equation}%
while the energy conservation equations for $\rho _{r}$ and $\rho _{v}$ can again
be written as in Eqs. (\ref{dr}) and (\ref{dv}). The energy density in the
non-interacting field, which we take to be separately conserved, is given by 
\begin{equation}
\dot{\rho}+3\gamma H\rho =0\,.  \label{dm}
\end{equation}%
At this point it is convenient to use the number of e-foldings, $N\equiv \ln a $, as a replacement for the time variable. We can then integrate Eq. (\ref{dm}) to find
\begin{equation}
\rho = \rho_0 e^{-3 \gamma N} \, ,
\end{equation}
where $\rho_0$ is a constant. The corresponding energy densities in the radiation and the vacuum can be found from Eqs. (\ref{bb}), (\ref{T}) and (\ref{fried}) to be
\begin{equation}
\rho_r = \frac{g_{\ast} H^4}{480 \pi^2} \left( 1+ \frac{H^{\prime}}{2 H} \right)^4
\end{equation}
and
\begin{equation}
\rho_v = \frac{3 H^2}{8 \pi} - \frac{g_{\ast} H^4}{480 \pi^2} \left( 1+ \frac{H^{\prime}}{2 H} \right)^4 - \rho_0 e^{-3 \gamma N}  \, ,
\end{equation}
where we have used a prime to denote a derivative with respect to $N$. Finally, differentiating Eq. (\ref{fried}), and using the conservation equations for $\rho_r$, $\rho_v$ and $\rho$ we obtain
\begin{equation}
H H^{\prime} = -\frac{4 \pi}{3} \left[ \frac{g_{\ast} H^4}{120 \pi^2} \left( 1+ \frac{H^{\prime}}{2 H} \right)^4 +3 \gamma \rho_0 e^{-3 \gamma N} \right] \, .
\end{equation}
This is a first-order ODE for $H$, as a function of $N$. Once we have $H=H(N)$, then the equations above give us $\rho_r$, $\rho_v$ and $\rho$ as functions of $N$, too.

\begin{figure}[t!]
\begin{center}
\includegraphics[scale=1.5]{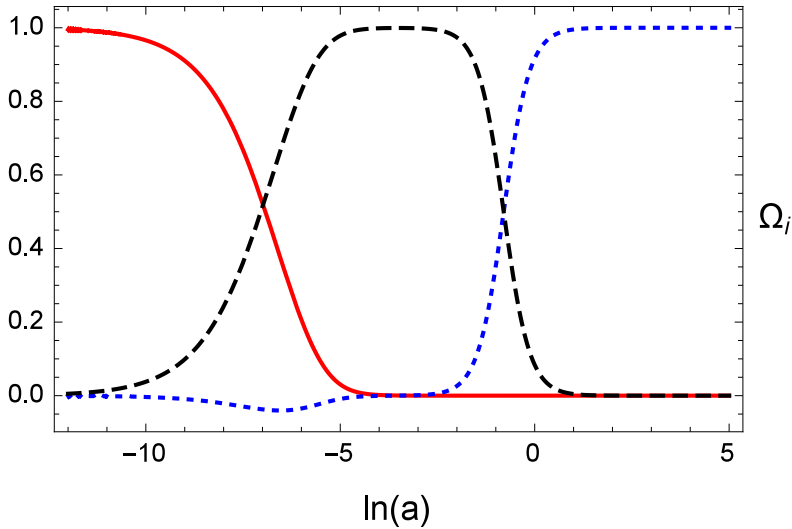}
\end{center}
\caption{The evolution of the density parameters for the radiation field (red solid line), the vacuum energy (blue dotted line) and the non-interacting fluid (black dashed line). To produce this plot we have taken the non-interacting fluid to be a pressure-less fluid of dust, such that $\gamma=1$.  We have also taken $g_{\ast}=2$, and set $H=0.1$ and $\rho=10^{-4}$ at $\ln a = 0$.}
\label{fig2}
\end{figure}

An example evolution of the density parameters of the radiation field, the vacuum field, and non-interacting fluid are shown in Fig. \ref{fig2}. To produce this plot we have taken the non-interacting fluid to be dust, so that $\gamma=1$. We have also chosen to consider the case $g_{\ast}=2$, and have set initial conditions so that $H=0.1$ and $\rho=10^{-4}$ at $\ln a =0$. For a finite period of time the energy density in the non-interacting dust dominates over the radiation and vacuum energies, and we have $a(t)\sim t^{2/3}$.  After this we have a period of intermediate inflation occurring, of the type given in Eq. (\ref{ii}). Before it we have radiation domination, as described in Eq. (\ref{rad}). During this early period of radiation domination, the vacuum energy in fact becomes negative, and starts to diverge. This change of sign does not appear in the absence of the non-interacting dust.

\begin{figure}[t!]
\begin{center}
\includegraphics[scale=1.5]{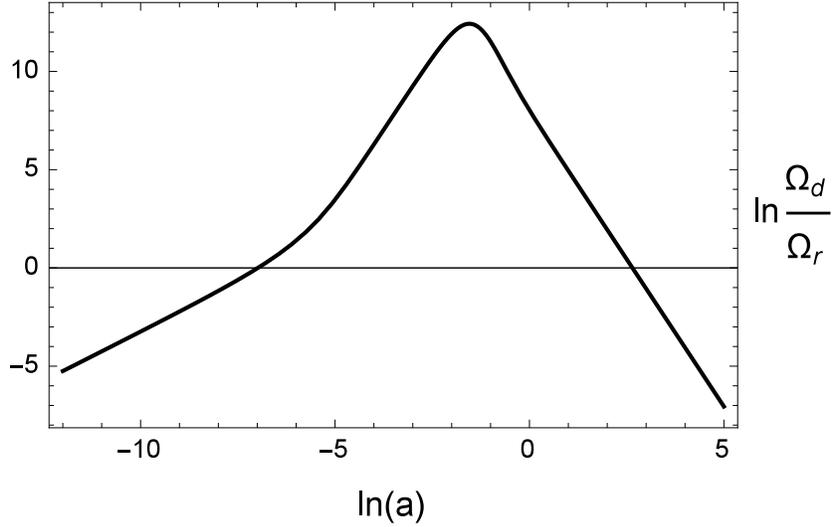}
\end{center}
\caption{The logarithm of the ratio in energy densities of dust and radiation, when $g_{\ast}=2$, and when $H=0.1$ and $\rho=10^{-4}$ at $\ln a=0$. The radiation field dominates over the dust at both late and early times.}
\label{fig3}
\end{figure}

At late times the energy density in radiation generically dominates over matter, in contrast to the usual
case in cosmological models with non-interacting radiation and dust \cite{rad}. In Fig. \ref{fig3} we plot the energy density in dust as a fraction of the energy density in radiation, again for $g_{\ast }=2$ and with $H=0.1$ and $\rho=10^{-4}$ at $\ln a =0$. It can be seen that there is a transient period when the dust dominates over the radiation, but that the radiation dominates over the dust both before and after this. The opposite result is true for $\gamma \leqslant 0$, in which case the non-interacting fluid dominates over radiation at late times, while being sub-dominant beforehand. If we had introduced an effective perfect `fluid' with $\gamma =2/3$, to mimic the presence of negative spatial curvature in the Friedmann equation, then the same general evolution occurs and the curvature `fluid' does not dominate at late times. This shows that flatness is approached at late times, just as in standard inflation.

We should point out at this stage that the black-body radiation that is created by the assumed thermal equilibrium with the vacuum is not the only radiation that one should expect in a realistic cosmology. For example, there is also a bath of radiation in the late Universe, which is very close to being a black-body, and that is at a much higher temperature than that of the apparent horizon. The synthesis of the light elements occurs during the epoch in which this fluid dominates over all other matter, and the energy that creates this additional radiation comes (originally) from the reheating process that occurs after inflation. A first approximation to the cosmological consequences of including an additional radiation field of this type, at a different temperature to the vacuum, could be studied by adding a non-interacting fluid, as described above. In this case the primordial synthesis of light elements should be expected to occur in exactly the same way that  it usually does, as the energy density of the vacuum is tiny compared to that of radiation during nucleosynthesis. 

In reality, of course, one would expect any additional fluid to also interact with the vacuum in some way, and perhaps stimulate an increase or decrease in the vacuum energy density by some small amount. A calculation to determine how this proceeds would require a knowledge of the non-equilibrium thermodynamics of the interaction. If this were known, and all of the carriers of entropy could be identified, then it would also be possible to investigate the stability of the assumed thermal equilibrium. We will leave a detailed study of these points for future work.

\section{Perturbations Generated During Inflation}

It is natural to consider the effects of the interaction introduced in
Section \ref{sec1} on the observables that result from thermal fluctuations
during inflation. To do this, we model the vacuum energy as a scalar field
with a self-interaction potential, $V(\phi )$, such that 
\begin{equation}
\rho _{\mathrm{v}}=\frac{1}{2}\dot{\phi}^{2}+V(\phi )\quad \mathrm{and}\quad
p_{\mathrm{v}}=\frac{1}{2}\dot{\phi}^{2}-V(\phi )\,.
\end{equation}%
The evolution we get in this case will be different to that obtained in the previous section, where we considered a fluid with equation of state $p_v = -\rho_v$, but will reduce to it in the appropriate limits. 

In terms of these new variables, the Friedmann and conservation equations
can be manipulated into the form 
\begin{eqnarray} \label{dH}
\dot{H}-\frac{4 g_{\ast } \pi^3 }{45 } T^{4}+3H^{2}-8\pi V &=&0 \\
\dot{\phi}^{2}+2V-\frac{3H^{2}}{4\pi }+\frac{g_{\ast } \pi ^2 }{15}T^{4}
&=&0\, , \label{dphi}
\end{eqnarray}%
with $T=T(H,\phi)$ given by the solution of
$
H T - g_{\ast} \pi^2 T^4/45 = 2 V - H^2 /4\pi \, ,
$
and where we have taken $H>0$ and $\dot{H}+2 H^2 >0$.
This simple system of equations represents a 2-dimensional dynamical system, the solutions of which can be found only after the form of the potential $V(\phi )$ is specified. In the absence of the terms involving $g_{\ast }$, these equations reduce to the usual ones for a scalar
field-filled Friedmann model.

Unlike in normal inflation, the existence of radiation should be expected to have an effect on the power spectrum of comoving curvature perturbations. This is because the thermal radiation should drive the evolution of the perturbations in $\phi$ through the existence of both noise and  dissipation terms. Such scenarios have been considered in the literature under the title ``warm inflation''. In this case the origin of the fluctuations in the CMB are not only from a quantum origin, but also from the thermal nature of the radiation.

If we differentiate Eq. (\ref{dphi}), and make use of Eq. (\ref{dH}), then we can write the Klein-Gordon equation for $\phi$ as
\begin{equation}
\ddot{\phi} + 3H \left( 1+ \Gamma \right) \dot{\phi} + \frac{dV}{d\phi} = 0 \, ,
\end{equation}
where we have included a dissipation term, with coefficient
\begin{equation} \label{gamma}
\Gamma = - \frac{8 g_* \pi^3 T^4}{45 H}  \frac{\left( H+\frac{\dot{T}}{T} \right) }{
  \left( \dot{H} +\frac{8 g_* \pi^3}{45} T^4 \right)} \, ,
\end{equation}
and where $T$ can be given in terms of $H$ using Eq. (\ref{T}).
The effect of $\Gamma$ on the spectrum of comoving curvature perturbations, $\mathcal{P_R}$, has been studied in a series of papers by Berera and collaborators \cite{warm1,warm2,warm3,warm4,warm5,warm6,warm7}.

If the noise source that the radiation creates is taken to be Markovian, then the spectrum of perturbations is found to be \cite{warm5,warm6,warm7}
\begin{equation}
\label{PR}
\mathcal{P_R} = \frac{H^4}{(2\pi)^2 \dot{\phi}^2}  \left[1+ \left( \frac{2 \pi T}{H} \right)
\frac{\Gamma}{\sqrt{1+\frac{4 \pi}{3} \Gamma}}  \right] \, .
%\mathcal{P_R} \simeq \frac{H^{4}}{(2\pi )^{2}\dot{\phi}^{2}} \left( 1+ \frac{\Gamma}{3H}  %\right)^{1/2}  \left( \frac{2 \pi T}{H} \right) \, .
\end{equation}
This expression reduces to the usual one when $\Gamma \rightarrow 0$. Generalised to include non-trivial distributions of inflaton particles \cite{warm5}, and noise sources for the radiation \cite{warm6}, have also recently been derived in the literature. 

The spectral index of the primordial curvature perturbations, $n_{s}$, is given in this case by the usual expression: 
\begin{equation}
n_{s}-1\equiv \frac{d\ln \mathcal{P}_{\mathcal{R}}}{d\ln k}\,,
\end{equation}%
where the right-hand side is to be evaluated at horizon crossing, when $k\simeq aH$. This spectral index can be calculated explicitly, to find a lengthy expression, using the equations above. Below we will calculate the leading-order contributions that occur during slow-roll inflation.

Likewise, the spectrum of tensor perturbations is found to be given by
\cite{warm4}
\begin{equation}
\mathcal{P}_{\mathrm{grav}} = \frac{4H^{2}}{\pi}  \, ,
\label{Pgrav}
\end{equation}
which can be used to write the  spectral index of tensor perturbations as
\begin{equation}
n_{T}-1\equiv \frac{d\ln \mathcal{P}_{\mathrm{grav}}}{d\ln k}
=\frac{2 \dot{H}}{(\dot{H}+ H^{2})} \, ,  \label{nt}
\end{equation}
where the derivative of $\ln \mathcal{P}_{\mathrm{grav}}$ has been evaluated at horizon crossing, to get the simple expression after the last equality. The spectral index of tensor perturbations can be seen to be unchanged by the presence of radiation. 

Lastly, Eqs. (\ref{PR}) and (\ref{Pgrav}) can also be used to define the tensor-to-scalar ratio:
\begin{eqnarray}
r \equiv \frac{A_T}{A_S} =4 \frac{\mathcal{P}_{\mathrm{grav}}}{\mathcal{P_R}}&=& 
-\frac{
16 \left( \frac{\dot{H}}{H^2} +\frac{g_* H^2}{90 \pi}  \right)
}{
\left[
1+ \left( \frac{2 \pi T}{H} \right)
\frac{\Gamma}{\sqrt{1+\frac{4 \pi}{3} \Gamma}}
   \right]
}
  \label{rdef} \,,
\end{eqnarray}%
where all roots are taken to be positive, and where we have used Eqs. (\ref{dH}) and (\ref{dphi}) to write this expression in terms of $H$ and its derivatives only. Here we have taken the amplitude of tensor perturbations to be $A_T=4 \mathcal{P}_{\mathrm{grav}}$, and the amplitude of scalar perturbations to be $A_S=\mathcal{P_R}$. Once more, this expression reduces to the usual one when $g_* \rightarrow 0$.

Recent observations imply that the scalar spectral index generated during inflation is given by \cite{planck} 
\begin{equation} \label{nsobs}
n_S-1 = -0.0365 \pm 0.0094, \,
\end{equation}
while the amplitude of curvature perturbations is inferred to be 
 \begin{equation} \label{asobs}
A_S = 2.19^{+0.12}_{-0.14} \times 10^{-9} \, ,
\end{equation}
and the tensor-to-scalar ratio is constrained by \cite{bicep} 
\begin{equation}
r \lesssim 0.2 \, .
\end{equation}
If we apply the latter two of these together, then Eq. (\ref{Pgrav}) can be seen to imply
\begin{equation}
\label{Hcon}
|H|\lesssim 10^{-5}\,.
\end{equation}%
This severely limits any effect that the radiation can have on the scalar spectral index, and the amplitude of scalar fluctuations. 

In fact, if we define slow-roll parameters by 
\begin{eqnarray}
\epsilon _{H} &\equiv &\frac{3\dot{\phi}^{2}}{2V+\dot{\phi}^{2}}=-\frac{(%
\dot{H}+\frac{8 g_{\ast } \pi^3 }{45} T^{4})}{(H^{2} -\frac{4 g_{\ast } \pi^{3}}{45} T^4)}  \label{e} 
%\end{eqnarray}%
%and
%\begin{eqnarray} 
\\
\eta _{H} &\equiv &-\frac{\ddot{\phi}}{H\dot{\phi}}=-\frac{(\ddot{H}+\frac{%
32 g_{\ast } \pi^{3}}{45} T^3 \dot{T})}{2H(\dot{H}+\frac{8 g_{\ast }\pi^{3}}{45} T^4)}%
\, ,  \label{n}
\end{eqnarray}%
then we can write our expressions for $A_S$ and $n_S$ as functions of $\epsilon_H$, $\eta_H$ and $H$ only. If we expand these first in $H$ (this quantity having been found to be small already), and then in $\epsilon_H$ and $\eta_H$, we find that the leading-order parts $A_S$ and $n_S$ are given by
\begin{equation} \label{nsapprox}
n_S-1 \simeq -4 \epsilon_H+2 \eta_H -\frac{g_{\ast} H^2}{45 \pi} \frac{(4\epsilon_H-\eta_H)}{\epsilon_H}
\end{equation}
and
\begin{equation}
\label{asapprox}
A_S \simeq -\frac{H^2}{\pi \epsilon_H} \left( 1+ \frac{g_{\ast} H^2}{90\pi \epsilon_H} \right) \, .
\end{equation}
The observational constraints from Eqs. (\ref{nsobs}) and (\ref{asobs}) then imply
\begin{equation}
\epsilon _{H}  \lesssim 0.1
\qquad {\rm and} \qquad
\eta _{H} \lesssim 0.1\, ,
\end{equation}
as long as $g_{\ast }\ll 10^{7}$. The observational constraints on $\epsilon _{H}$ and $\eta _{H}$ are therefore unchanged from their usual values. The expressions for the spectral indices of scalar and tensor fluctuations, as well as their amplitudes, are effectively given by the usual expressions in terms of the slow-roll parameters, with leading-order corrections as given in Eqs. (\ref{nsapprox}) and (\ref{asapprox}).

Given these constraints, one might naively expect that the level of non-gaussianity should be the same as that in standard slow-roll inflationary models, where $f_{NL}\sim O(\epsilon _{H}$
and $\eta _{H})\sim 10^{-2}$, \cite{cheng}.  Detailed calculations, however, show that the amplitude of non-gaussianity are strongly dependant on the parameters involved in the interactions of the warm inflationary model, and hence on the microscopic physics and dynamics \cite{ng}. For models with weak interactions, as suggested by the observational constraints above, the shape of the bispectrum is found to be close to equilateral.

\section{Discussion}

In this paper, we have studied the cosmological consequences of the vacuum
being in thermal equilibrium with a bath of black-body radiation, as suggested by Paddy in \cite{pad}. In this
situation, energy is exchanged between the vacuum and the radiation. In the
absence of other matter fields, the assumption of a vacuum equation of state 
$p_{v}=-\rho _{v}$, and a temperature corresponding to the surface gravity of the cosmological horizon, results at late times in intermediate inflation with $H\propto t^{-1/3}$, and the introduction of an
initial curvature singularity.

We also calculated the evolution of such a universe when it
contains a non-interacting barotropic perfect fluid, in additional to the
interacting radiation and vacuum energy. We found that, as long as the
non-interacting fluid has an equation of state $p>-\rho $, it is dominated
at both late and early times by the radiation. We also find that it is
possible for the non-interacting fluid to dominate for a finite period at
intermediate times, and that during this time the energy density in the vacuum can change sign from negative to positive.

We then proceeded to study the observational consequences of this energy exchange when
the vacuum is treated as a slowly-rolling minimally-coupled scalar field
with a self-interaction potential. We find that observational constraints on
the amplitude of scalar and tensor perturbations, and the spectral index of
primordial curvature perturbations, result in expressions that are very
close to the usual ones, written in terms of the slow-roll parameters. There
are therefore no strong observational constraints to distinguish this
scenario from a standard slow-roll inflation.

While the generic end-state of these models is intermediate inflation
driven by the vacuum energy, we find that the generic initial state is
a radiation dominated universe in which all energy densities diverge. The occurrence of an early universe with a large negative vacuum energy, that can evolve into a late universe with positive vacuum energy, is an intriguing consequence of this scenario, and would appear to be consistent with the
picture of the negative Planck-sized vacuum energy that is generically expected to result from the lowest-order super-gravity terms in string and M-theories \cite{ads}. We leave further study of this feature for future work.

\section*{Acknowledgements}

We are grateful to J.~Lidsey, A.~Linde and D.~Mulryne for helpful discussions. TC and JDB are both supported by the STFC.

\end{document}